\documentclass[nofootinbib, superscriptaddress, reprint, amsmath,amssymb, aps]{revtex4-1}
\usepackage{graphicx}
\usepackage{bm}
\graphicspath{ {./images/} }

\usepackage[dvipsnames]{xcolor}
\usepackage{footnote}
\usepackage{enumitem}
\usepackage{qcircuit}
\usepackage{float}
\usepackage{mathtools}
\usepackage{amsmath}
\usepackage[colorlinks]{hyperref}

\usepackage{verbatim}

\begin{document}

\title{Efficient quantum circuits for quantum computational chemistry}

\author{Yordan S. Yordanov}
\affiliation{Cavendish Laboratory, Department of Physics, University of Cambridge, Cambridge CB3 0HE, United Kingdom}
\author{David R. M. Arvidsson-Shukur}
\affiliation{Cavendish Laboratory, Department of Physics, University of Cambridge, Cambridge CB3 0HE, United Kingdom}
\affiliation{Research Laboratory of Electronics, Massachusetts Institute of Technology, Cambridge, Massachusetts 02139, USA}
\author{Crispin H. W. Barnes}
\affiliation{Cavendish Laboratory, Department of Physics, University of Cambridge, Cambridge CB3 0HE, United Kingdom}

\begin{abstract}
Molecular quantum simulations with the  variational quantum eigensolver (VQE) rely on ansatz states that approximate the molecular ground states. These ansatz states are generally defined by parametrized fermionic excitation operators and an initial reference state. Efficient ways to perform fermionic excitations are vital for the realization of the VQE on noisy intermediate-scale quantum computers. 
Here, we address this issue by first demonstrating circuits that perform qubit excitations, excitations that do not account for fermionic anticommutation relations. We then extend the functionality of these circuits to perform fermionic excitations.
Compared to circuits constructed with the standard use of ``$CNOT$ staircases'', our circuits offer a linear reduction in the number of $CNOT$ gates, by a factor of $2$ and $8$ per single and double excitation, respectively.
Our results reduce the requirements for near-term realizations of quantum molecular simulations.
\end{abstract}

\maketitle


\section{ Introduction}\label{Intro}

It is anticipated that quantum computers will be able to simulate quantum systems exponentially faster than classical computers \cite{QC_gen_1, QC_gen_2, uni_q_sim}.
A promising method to perform this task on emerging noisy intermediate-scale quantum (NISQ) computers \cite{NISQ,google_supreme} is the variational-quantum-eigensolver (VQE) algorithm \cite{UCCSD_0, vqe_2, VQE_accelerated, vqe_google_hf}. 
The VQE is a  hybrid classical-quantum variational algorithm that determines the eigenvalues of a Hamiltonian operator by minimizing its expectation value, with respect to a parametrized ansatz state. The VQE can be used to solve the electronic structure problem, determining the ground-state energy eigenvalue of a molecular Hamiltonian \cite{vqe_general, CCSD}. 
Compared to other purely quantum algorithms for eigenvalue-determination, e.g., the quantum phase-estimation algorithm \cite{QC_QI,QPE}, the VQE uses shallower quantum circuits and is more error tolerant \cite{UCCSD_0}, at the expense of requiring a greater number of measurements and more classical processing.

Two major challenges in the practical realization of the VQE on NISQ computers, are the design of ansatz states, and the construction of efficient circuits to create these states.
Most ansatz states, currently considered by the scientific community, correspond to applying a series of fermionic unitary evolutions to an initial reference state, e.g., the Hartree-Fock state. These fermionic unitary evolutions, which we will refer to as ``fermionic excitations'', are exponentials of parametrized single- and double-fermionic-excitation operators. Examples of such fermionic ansatz states include unitary coupled cluster (UCC) ansatz states \cite{UCCSD_0, UCCSD, UCCSD_2, k-upCCGSD}, and states constructed by iterative-VQE algorithms, e.g. the adaptive derivative-assembled pseudo-trotter (ADAPT) VQE \cite{vqe_adapt, benchmark_adapt_vqe}.
The standard circuits  for fermionic excitations \cite{vqe_general, UCCSD, vqe_circs, staircase_example_q} use ``$CNOT$ staircases'' (see Sec. \ref{sec:stair_case}) inefficiently, in terms of $CNOT$ gates. The ability to implement a sufficient number of entangling gates with high enough fidelity is the current bottleneck in NISQ computing \cite{NISQ, CNOT_fidelity, CNOT_fidelity_2}. 
Therefore, more efficient fermionic-excitation  circuits will improve the prospects of near-term  realization of the VQE on NISQ computers.

In this work, we demonstrate a methodology to construct circuits, optimized in the number of $CNOT$ gates and in the $CNOT$ circuit depth, that perform single and double fermionic excitations. 
First, we construct simpler excitations that do not account for fermionic anticommutation relations. We call these simplified excitations ``qubit excitations''.
Single qubit excitations can be implemented by an exchange-interaction circuit \cite{part_hole,exchange_ansatz, entanglement_generation, VQE_hard_symmetry_preserve}. 
Second, we use this exchange-interaction circuit as a subcircuit to construct a double qubit excitation circuit. 
Finally, we modify these qubit excitations circuits, to account for fermionic anticommutation relations.
In this way we construct circuits that offer a linear reduction in the number of $CNOT$ gates by a factor of $2$ and $8$ per single and double fermionic excitation, respectively, compared to circuits constructed entirely with $CNOT$ staircases.

The paper is organised as follows: In
Sec. \ref{sec:stair_case}, we review the standard way of using  $CNOT$ staircases to construct fermionic-excitation circuits. In Sec. \ref{sec:ncU1}, we outline a method, which is used throughout the paper, to implement multi-qubit-controlled rotations. In Sec. \ref{sec:qubit_exc}, we design circuits that perform single and double qubit excitations. In Sec. \ref{sec:f_exc}, we utilise our qubit excitations to construct circuits that perform fermionic excitations. We summarize our results in Sec. \ref{summary}.

\section{Theory}\label{sec:theory}

Throughout this paper we assume the Jordan-Wigner encoding \cite{vqe_general, qubit_mapping}, where the occupancy of the $i^{\mathrm{th}}$ molecular spin orbital  is represented by the state of  qubit $q_i$. We denote spin orbitals with indices $i,j,k,l$, ordered as $i<j<k<l$. 

\subsection{Fermionic excitations}\label{sec:stair_case}

In this section, we provide a brief description of what a fermionic excitation is, within the context of molecular VQE simulations. We also  review the standard method to construct circuits that perform fermionic excitations \cite{vqe_general, UCCSD, vqe_circs, staircase_example_q}.

We use the term fermionic excitation to refer to an exponential of $\theta$-parametrized fermionic excitation operators.
Single and double fermionic excitation operators are defined, respectively, by the skew-Hermitian operators
\begin{equation}\label{eq:s_f_exc_op}
T^k_i(\theta) \equiv \theta( a^\dagger_k a_i - a^\dagger_i a_k)\;\;\mathrm{and}
\end{equation}
\begin{equation}\label{eq:d_f_exc_op}
T_{ij}^{kl}(\theta) \equiv \theta( a^\dagger_k a^\dagger_l a_i a_j - a^\dagger_i a^\dagger_j a_k a_l),
\end{equation}
where  $a_i^\dagger $ and $a_i$ denote fermionic creation and annihilation operators, respectively.
Single and double fermionic excitations are expressed, respectively, by the unitary operators
\begin{equation}\label{eq:s_f_exc_aa}
U_{ki}^{\mathrm{(sf)}}(\theta) = e^{ T^{k}_{i}(\theta) }\;\;\mathrm{and}
\end{equation}
\begin{equation}\label{eq:d_f_exc_aa}
U_{klij}^{\mathrm{(df)}}(\theta) = e^{ T^{kl}_{ij}(\theta) }.
\end{equation}
The operators $a_i$ and $a_i^\dagger $ obey the anticommutation relations
\begin{equation}
\label{eq:fermAnti}
\{a_i,a^\dagger_j\} = \delta_{i,j} \; \; \mathrm{and} \; \; \{a_i, a_j\} = \{a_i^\dagger, a^\dagger_j\} = 0.
\end{equation}
Within the Jordan-Wigner encoding, $a$ and $a^\dagger$ can be written in terms of quantum-gate operators as
\begin{equation}\label{eq:a}
a_i =  Q_i \prod_{r=0}^{i-1} Z_r=\frac{1}{2}(X_i+iY_i)\prod_{r=0}^{i-1} Z_r \;\;\mathrm{and}
\end{equation}
\begin{equation}\label{eq:a*}
a_i^\dagger  = Q_i^\dagger \prod_{r=0}^{i-1} Z_r = \frac{1}{2}(X_i-iY_i)\prod_{r=0}^{i-1} Z_r,
\end{equation}
where the qubit creation and annihilation operators are
\begin{equation}\label{eq:Q_Q_0}
Q_i^\dagger \equiv \frac{1}{2}(X_i - iY_i)  \; \; \mathrm{and} \; \;  Q_i \equiv \frac{1}{2}(X_i + iY_i) ,
\end{equation}
respectively. $Q_i$ and $Q_i^{\dagger}$ (discussed further in Sec. \ref{sec:qubit_exc})  act to change the occupancy of orbital $i$. The product of Pauli-$z$ operators, $\prod_{r=0}^{i-1} Z_r$, acts as an exchange-phase factor, accounting for the anti-commutation relations of $a$ and $a^\dagger$.

\begin{figure}[h]
\ \ \ \ \ \Qcircuit @C=0.9em @R=0.7em {
q_k && \gate{R_x(\frac{\pi}{2})} & \ctrl{1} & \qw & \qw &  \qw &\qw & \qw &\ctrl{1}& \gate{R_x(-\frac{\pi}{2})} & \qw \\
q_{k-1} && \qw & \targ & \ctrl{1} & \qw & \qw & \qw &\ctrl{1} &\targ & \qw & \qw\\
\vdots &&&& \vdots &&&& \vdots &&& \\
q_{i+1} && \qw & \qw & \targ &  \ctrl{1} & \qw & \ctrl{1}  & \targ & \qw & \qw & \qw\\
q_{i} && \gate{H}  &\qw & \qw & \targ & \gate{R_z(\theta)} & \targ & \qw & \qw & \gate{H} & \qw\\
}
\caption{A circuit implementing the exponential $\exp\big[-i\theta X_i Y_{k} \prod_{r=i+1}^{k-1} Z_r\big]$.}
\label{fig:stair_case}
\end{figure}
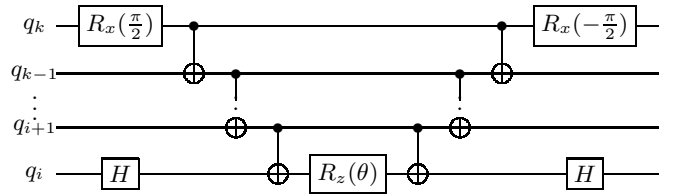

Using Eqs. (\ref{eq:a}) and (\ref{eq:a*}), a single fermionic excitation [Eq. \eqref{eq:s_f_exc_aa}] can be re-expressed in terms of quantum-gate operators:
\begin{align}
 U_{ki}^{\mathrm{(sf)}}(\theta) = &\exp\Big[-i\frac{\theta}{2}(X_i Y_k - Y_i X_k)\prod_{r=i+1}^{k-1}Z_r\Big]  \\   = &
\exp\Big[-i\frac{\theta}{2}X_i Y_k\prod_{r=i+1}^{k-1}Z_r\Big] \nonumber \\ &\times \exp\Big[i\frac{\theta}{2}Y_i X_k\prod_{r=i+1}^{k-1}Z_r\Big]  \label{eq:s_f_exc_0}  .
\end{align}
The exponential $\exp\big[-i\frac{\theta}{2}X_i Y_k\prod_{r=i+1}^{k-1}Z_r\big]$  in Eq. (\ref{eq:s_f_exc_0}), can be implemented by the circuit in Fig. \ref{fig:stair_case}. The two $CNOT$ ``staircases'' together with the $R_z(\theta)$ rotation in between them (Fig. \ref{fig:stair_case}), are referred to as a ``$CNOT$-staircase construction''. This construction implements $U_{stair}\equiv \exp\big[i\theta \prod_{r=i}^{k} Z_r\big]$.
The $H$ and $R_x(\pm \frac{\pi}{2})$ gates, on both sides of the $CNOT$ staircase construction in Fig. \ref{fig:stair_case}  act to transform the $Z_i$ and the $Z_k$  operators, in $U_{stair}$, to $X_i$ or $Y_k$ operators, respectively.
Similarly, by sandwiching a $CNOT$ staircase construction between single-qubit rotations that transform individual Pauli-$z$ operators to Pauli-$x$ or Pauli-$y$ operators, any exponential of a product of Pauli operators can be constructed.
Hence, the single fermionic excitation [Eq. \eqref{eq:s_f_exc_0}] can be implemented by a circuit that contains $2$ $CNOT$ staircase constructions. 
The full circuit for a single fermionic excitation, constructed using the aforementioned method, is included in appendix \ref{app:f_circs} Fig. \ref{fig:s_f_exc}.

A double fermionic excitation [Eq. \eqref{eq:d_f_exc_aa}] can be re-expressed in terms of quantum-gate operators:
\begin{multline}\label{eq:d_f_exc}
\ \ \ \ \ \ \ \ \ \ \ U_{klij}^{\mathrm{(df)}}(\theta) = \exp\Big[-i \frac{\theta}{8} (X_i Y_j X_k X_l + Y_i X_j X_k X_l \\  \\+ Y_i Y_j Y_k X_l  +  Y_i Y_j X_k Y_l  - X_i X_j Y_k X_l  - X_i X_j X_k Y_l \\ - Y_i X_j Y_k Y_l - X_i Y_j Y_k Y_l ) \prod_{r=i+1}^{j-1} Z_r\prod_{r'=k+1}^{l-1} Z_{r'}\Big]. \ \ \ \
\end{multline}
$U_{klij}^{\mathrm{(df)}}(\theta)$ can be implemented by  $8$ $CNOT$-staircase constructions.
The full circuit for a double fermionic excitation, constructed using the aforementioned method, is included in appendix \ref{app:f_circs} Fig. \ref{fig:d_f_exc}.

\subsection{Multi-qubit-controlled rotations}\label{sec:ncU1}

In this section, we outline a method, used in \cite{POVM}, to construct circuits that implement multi-qubit-controlled rotations. The circuits use no ancilla qubits.
Later, we use this method to construct circuits for qubit excitations (see Sec. \ref{sec:qubit_exc}).

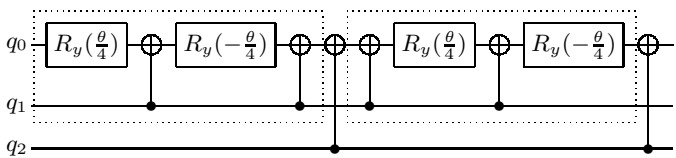
\begin{figure}[H]
\Qcircuit @C=0.6em @R=1.5em {
&q_0 && \gate{R_y(\frac{\theta}{4})}  & \targ{} & \gate{R_y(-\frac{\theta}{4})}  & \targ{} & \targ{} &  \targ{} & \gate{R_y(\frac{\theta}{4})}  & \targ{} & \gate{R_y(-\frac{\theta}{4})}  & \targ{} & \qw  \\
&q_1 && \qw  & \ctrl{-1} & \qw & \ctrl{-1} & \qw & \ctrl{-1} & \qw  & \ctrl{-1} & \qw  & \qw  & \qw\\
&q_2 && \qw & \qw & \qw & \qw & \ctrl{-2} & \qw & \qw & \qw & \qw &  \ctrl{-2} & \qw\gategroup{1}{4}{2}{7}{1.0em}{..} \gategroup{1}{9}{2}{12}{1.0em}{..}
}
\caption{A circuit implementing $R_y(\theta, \{q_1,q_2\}, q_0)$. The first half-way rotation $R_y(\frac{\theta}{2}, \{q_1\}, q_0)$ (left dotted rectangle) is implement as in Eq. \eqref{cU_1}, and the second half-way rotation $R_y(-\frac{\theta}{2}, \{q_1\}, q_0)$ ( right dotted rectangle) as in Eq. \eqref{cU_2}. In this way, the two middle $CNOT$s between qubits $q_0$ and $q_1$ can be cancelled.}
\label{fig:ccU1}
\end{figure}

We denote a $R_y(\theta)$ rotation on a target qubit $q_0$, controlled by the $\{q_1..q_m\}$ qubits being in the state $|1..1\rangle$, as $R_y\big(\theta, \{q_1..q_m\}, q_0\big)$. We then write an $m$-qubit-controlled rotation by decomposing it into two opposite half-way rotations, controlled by $(m-1)$ qubits:
\begin{align}\label{cU_1}
R_y\big(\theta, \{q_1..q_m\}, q_0\big) = \nonumber  \\ CNOT(q_1, q_0) R_y\big(-\frac{\theta}{2}, \{q_2..q_m\}, q_0\big) \nonumber  \\ \times  CNOT(q_1, q_0)R_y\big(\frac{\theta}{2}, \{q_2..q_m\},q_0\big),
\end{align}
or equivalently, as
\begin{align}\label{cU_2}
 R_y\big(\theta, \{q_1..q_m\}, q_0\big) =  \nonumber \\ R_y\big(\frac{\theta}{2}, \{q_2..q_m\}, q_0\big) CNOT(q_1, q_0) \nonumber \\ \times R_y\big(-\frac{\theta}{2}, \{q_2..q_m\}, q_0\big)CNOT(q_1, q_0).
\end{align}
By decomposing the controlled rotations further, the overall operation is brought down to $CNOT$s and single-qubit rotations.
Implementing directly Eqs. (\ref{cU_1}) or (\ref{cU_2}), results in a circuit with $(2^{m+1}-2)$ $CNOT$s. However, for $m>2$, Eqs. (\ref{cU_1}) and (\ref{cU_2}) can be combined alternately to cancel adjacent $CNOT$s (see Fig. \ref{fig:ccU1}), and obtain a circuit with $(2^{m-1}+2)$ $CNOT$s. $R_z$- and $R_x$-controlled rotations can be obtained by additional single-qubit rotations on the target qubit.
Equivalently, $CZ$ (controlled-phase) gates can be used instead of $CNOT$ gates in Eqs. (\ref{cU_1}) and (\ref{cU_2}); in some scenarios this can be used to cancel adjacent two-qubit gates (see Sec. \ref{sec:qubit_exc}).

\section{Results}\label{sec:results}

In this section we present our main results. We begin by defining qubit excitations. Then, we construct circuits to perform single and double qubit excitations. Last, we  modify these circuits to perform single and double fermionic excitations instead.

\subsection{Qubit excitations}\label{sec:qubit_exc}

We use the term qubit excitation to describe the unitary evolution of an exponential of a parametrized qubit  excitation operator. Single- and double-qubit-excitation operators are  generated by the qubit annihilation and creation operators, $Q$ and $Q^\dagger$ [Eq. (\ref{eq:Q_Q_0})], and are given by the skew-Hermitian operators
\begin{equation}\label{eq:s_q_exc_op}
\tilde{T}_i^k(\theta) \equiv \theta(Q^\dagger_k Q_i - Q^\dagger_i Q_k)\;\;\mathrm{and}
\end{equation}
\begin{equation}\label{eq:d_q_exc_op}
\tilde{T}^{kl}_{ij}(\theta) \equiv \theta(Q^\dagger_k Q^\dagger_l Q_i Q_j - Q^\dagger_i Q^\dagger_j Q_k Q_l).
\end{equation}
The operators $Q$ and $Q^\dagger$ satisfy the relations 
\begin{multline}\label{eq:Q_Q}
\ \ \ \{Q_i, Q_i^\dagger\} = I, \ \ [Q_i, Q^\dagger_j] = 0 \ \text{ if } \ i \neq j , \ \ \mathrm{and}\\ \ \ \ \  \  [Q_i, Q_j] = [Q_i^\dagger, Q^\dagger_j] = 0 \text{ for all }i,j.   \ \ \ \ \ \ \ \ \ \ \ \ \ \ \ 
\end{multline}
These relations are neither bosonic nor fermionic; some authors have referred to them as parafermionic \cite{parafermions}.

Single and double qubit excitations are expressed, respectively, by the unitaries 
\begin{equation}\label{eq:s_q_exc_QQ}
U_{ki}^{\mathrm{(sq)}}(\theta) = e^{\tilde{T}_i^k(\theta)} \;\;\mathrm{and}
\end{equation}
\begin{equation}\label{eq:d_q_exc_QQ}
 \ U_{klij}^{\mathrm{(dq)}}(\theta)  = e^{\tilde{T}^{kl}_{ij}(\theta)} .
\end{equation}
These unitary operations are similar to the single and double fermionic excitations [Eqs. \eqref{eq:s_f_exc_aa} and \eqref{eq:d_f_exc_aa}], apart from the absence of the exponentials of products of Pauli-$z$ operators that account for the fermionic anticommutation relations [Eq. \eqref{eq:fermAnti}].
Previously, these operations have been considered  in the context of VQE-ansatz construction \cite{UCCSD_qubit, Eff_d_q_exc}.

\subsubsection{Circuit for single qubit excitations}\label{sec:s_q_exc_QQ}

\begin{figure}[H]
\Qcircuit @C=0.75em @R=0.75em {
 \\
& \targ & \ctrl{1} &\qw & && \gate{R_y(-\frac{\pi}{2})} & \gate{R_z(-\frac{\pi}{2})} & \targ{} & \gate{R_z(\frac{\pi}{2})} & \gate{R_y(\frac{\pi}{2})} & \qw &  \\
&&&& =&&&&&&&&\\
& \ctrl{-2}  &\ctrl{-2} & \qw &&& \qw & \gate{R_z(\frac{\pi}{2})} &  \ctrl{-2} & \qw & \qw & \qw & 
}
\caption{A circuit identity for an operation equivalent to a $CNOT$ followed by a $CZ$.}
\label{fig:c_zx}
\end{figure}
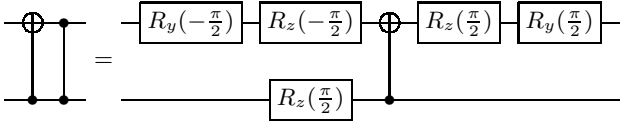

A single qubit excitation [Eq. (\ref{eq:s_q_exc_QQ})] can be re-expressed in terms of quantum-gate operators:
\begin{equation}\label{eq:s_exch}
U_{ki}^{\mathrm{(sq)}}(\theta) = \exp\Big[-i\frac{\theta}{2}(X_i Y_k - Y_i X_k)\Big] .
\end{equation}
This unitary is equivalent to an evolution under the exchange interaction, which can be performed by the circuit in Fig. \ref{fig:s_exchange}. 
We implement the controlled $R_y(\theta)$ rotation in the circuit in Fig. \ref{fig:s_exchange}a as in Eq. (\ref{cU_2}), and apply the circuit identity in Fig. \ref{fig:c_zx} to cancel a $CNOT$. In this way we obtain the explicit circuit, for single qubit excitation, shown in Fig. \ref{fig:s_exchange}b.\footnote{The  circuit in Fig. \ref{fig:s_exchange}b is not optimal in the number of $CNOT$ gates. A single qubit excitation, as defined in Eq. (\ref{eq:s_exch}), can be implemented by a circuit with only two $CNOT$s. We use the circuit in Fig. \ref{fig:s_exchange}b because it is used to construct the circuits in Figs. \ref{fig:d_q_exc}, \ref{fig:s_fermi} and \ref{fig:d_fermi} in an optimal way.}
 
\begin{figure}[H]
\Qcircuit @C=0.7em @R=1.0em {
\textbf{a)} &&&&&&&&&&&&& q_k && \ctrl{2}  & \gate{R_y(\theta)} & \ctrl{2} & \qw &\\
&&&&&&&&&&&&&&&&&&&=&&&&&\\
&&&&&&&&&&&&& q_i && \targ{} & \ctrl{-2} & \targ{} & \qw &\\
\\
}
\Qcircuit @C=0.5em @R=1.0em {
\textbf{b)}&&& q_k \ \ &&\qw&\gate{R_z(\frac{\pi}{2})} & \ctrl{2} &\gate{R_y(\frac{\theta}{2})}  & \ctrl{2} & \gate{R_y(-\frac{\theta}{2})}  & \ctrl{2} & \qw &\\
&&&&&&&&&&&&&&\\
&&& q_i \ \ && \gate{R_y(-\frac{\pi}{2})} & \gate{R_z(-\frac{\pi}{2})}  & \targ &\gate{R_z(-\frac{\pi}{2})}  & \targ & \gate{H} & \targ{} & \qw  
}
\caption{\textbf{a)} A circuit performing an exchange-interaction operation equivalent to the single qubit excitation in Eq. (\ref{eq:s_exch}). \textbf{b)} An explicit circuit for \textbf{a}) obtained by implementing the controlled-$R_y(\theta)$ rotation as in Eq. (\ref{cU_2}), and using the circuit identity in Fig. \ref{fig:c_zx}.}
\label{fig:s_exchange}
\end{figure}
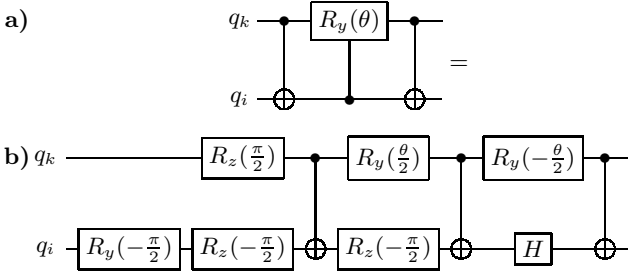

\subsubsection{Circuit for double qubit excitations}\label{sec:d_q_exc_QQ}

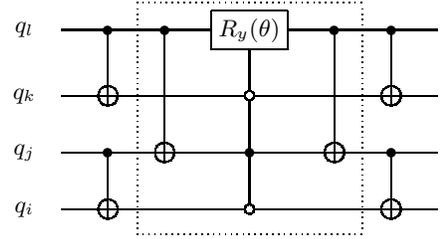
\begin{figure}[h]
\ \ \ \ \ \ \ \ \ \ \ \ \ \ \Qcircuit @C=1.5em @R=1.5em {
q_l && \ctrl{1} & \ctrl{2} & \gate{R_y(\theta)} & \ctrl{2} &  \ctrl{1} & \qw &\\
q_k && \targ{}  & \qw & \ctrlo{-1} & \qw &  \targ{} & \qw &\\
q_j  && \ctrl{1} &  \targ{} & \ctrl{-1} & \targ{} & \ctrl{1} & \qw &\\
q_i && \targ{} & \qw & \ctrlo{-1} & \qw & \targ{} & \qw \gategroup{1}{4}{4}{6}{2.0em}{..}
}
\caption{A circuit performing a double qubit excitation [Eq. (\ref{eq:d_q_exc})]. The explicit circuit is given in Fig. \ref{fig:d_q_exc_full}.}
\label{fig:d_q_exc}
\end{figure}

A double qubit excitation [Eq. (\ref{eq:d_q_exc})] can be re-expressed in terms of quantum-gate operators:
\begin{multline}\label{eq:d_q_exc}
\ \ \ \ \ \ \ \ \ \ \ U_{klij}^{\mathrm{(dq)}}(\theta) = \exp\Big[-i \frac{\theta}{8} (X_i Y_j X_k X_l + Y_i X_j X_k X_l  \\+ Y_i Y_j Y_k X_l  +  Y_i Y_j X_k Y_l  - X_i X_j Y_k X_l  - X_i X_j X_k Y_l \\ - Y_i X_j Y_k Y_l - X_i Y_j Y_k Y_l )\Big]. \ \ \ \ \ \ \ \ \ \ \ \ \ \ \ \ \ \ \ \ \ \ \ \ \ \ 
\end{multline}
Equation \eqref{eq:d_q_exc} is a unitary operation that continuously exchanges the $|1_i1_j0_k0_l\rangle$ and $|0_i0_j1_k1_l\rangle$ states, but does not alter other states.
To implement this operation, we  use a similar circuit to the one for the single qubit excitation. 
However, to ensure that the operation exchanges only the states $|1_i1_j0_k0_l\rangle$ and $|0_i0_j1_k1_l\rangle$, it must act nontrivially only if the parities of the qubit pairs $\{q_k q_l\}$ and $\{q_i q_j\}$ are even. 
To perform this kind of parity-controlled exchange operation, we use the circuit in Fig. \ref{fig:d_q_exc}. 
The first two $CNOT$s, between qubits $q_k$ and $q_l$, and qubits $q_i$ and $q_j$, encode the parities of the two respective qubit pairs on qubits $q_k$ and $q_i$, respectively.
Then qubits $q_k$ and $q_i$ are used as control qubits for a controlled exchange operation, the dotted rectangle in Fig. \ref{fig:d_q_exc}, between qubits $q_l$ and $q_j$. 
The last two $CNOT$s between qubits $q_k$ and $q_l$, and qubits $q_i$ and $q_j$, respectively, reverse the parity-encoding action of the first two $CNOT$s.

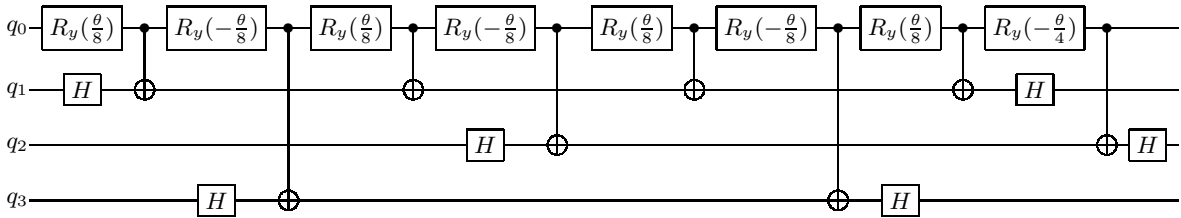
\begin{figure}[H]
\ \ \ \ \ \ \ \ \Qcircuit @C=0.5em @R=1.0em {
\\
q_0 && \gate{R_y(\frac{\theta}{8})}  & \ctrl{1} & \gate{R_y(-\frac{\theta}{8})}  & \ctrl{3} & \gate{R_y(\frac{\theta}{8})}  & \ctrl{1} & \gate{R_y(-\frac{\theta}{8})}  & \ctrl{2} & \qw  & \gate{R_y(\frac{\theta}{8})}  & \ctrl{1} & \gate{R_y(-\frac{\theta}{8})}  & \ctrl{3} & \gate{R_y(\frac{\theta}{8})}  & \ctrl{1} & \gate{R_y(-\frac{\theta}{4})}  & \ctrl{2} & \qw  & \qw  \\
q_1 && \gate{H}  & \targ & \qw  & \qw & \qw  & \targ & \qw  & \qw  & \qw & \qw  & \targ & \qw  & \qw & \qw  & \targ & \gate{H}  & \qw  & \qw & \qw   \\
q_2 && \qw & \qw & \qw & \qw & \qw & \qw & \gate{H} &  \targ & \qw & \qw & \qw & \qw & \qw & \qw & \qw & \qw &   \targ &  \gate{H} &  \qw \\
q_3 && \qw & \qw & \gate{H} & \targ & \qw & \qw & \qw &  \qw & \qw & \qw & \qw & \qw & \targ & \gate{H} & \qw & \qw & \qw & \qw & \qw  \\
\ \
}
\caption{An explicit circuit implementing the controlled rotation $R_y(\theta, \{q_1,q_2,q_3\}, q_0)$. The circuit is obtained with the method described in Sec. \ref{sec:ncU1}.}
\label{fig:cccU1}
\end{figure}

\begin{figure}[H]
\Qcircuit @C=0.3em @R=1.0em {
&&& \\ 
q_l \ && \ctrl{1} & \qw & \ctrl{2} & \gate{R_y(\frac{\theta}{8})} & \ctrl{1} & \gate{R_y(-\frac{\theta}{8})} & \ctrl{3} & \gate{R_y(\frac{\theta}{8})} & \ctrl{1} & \gate{R_y(-\frac{\theta}{8})} & \ctrl{2} & \gate{R_y(\frac{\theta}{8})} & \ctrl{1} & \gate{R_y(-\frac{\theta}{8})} & \ctrl{3} & \gate{R_y(\frac{\theta}{8})} & \ctrl{1} & \gate{R_y(-\frac{\theta}{8})} & \ctrl{2} &  \gate{R_z(\frac{\pi}{2})} & \qw & \ctrl{1} & \qw \\
q_k \ && \targ{} & \gate{X} & \qw & \gate{H} & \targ & \qw & \qw & \qw & \targ & \qw & \qw & \qw & \targ & \qw & \qw & \qw & \targ & \gate{H} & \qw & \qw & \gate{X} & \targ & \qw & \\
q_j \ && \ctrl{1} & \qw & \targ{} & \qw & \qw  & \qw & \qw & \qw & \qw & \gate{H} & \targ & \qw & \qw  & \qw & \qw & \qw & \qw & \gate{R_z(-\frac{\pi}{2})} & \targ & \gate{R_z(-\frac{\pi}{2})}& \gate{R_y(-\frac{\pi}{2})}  & \ctrl{1} & \qw & \\
q_i \ && \targ{} & \gate{X} & \qw & \qw & \qw & \gate{H} & \targ & \qw & \qw & \qw & \qw & \qw & \qw & \qw & \targ & \gate{H} & \qw & \qw & \qw  & \qw &\gate{X} & \targ & \qw &  \\
}
\caption{An explicit circuit implementing a double qubit excitation [Eq. (\ref{eq:d_q_exc})].}
\label{fig:d_q_exc_full}
\end{figure}
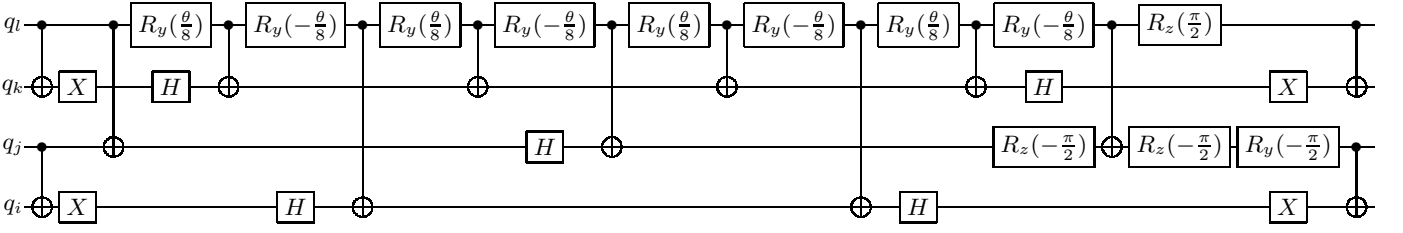

We implement the controlled $R_y(\theta)$ rotation in Fig. \ref{fig:d_q_exc} with the circuit in Fig. \ref{fig:cccU1}, and use the circuit identity in Fig. \ref{fig:c_zx} to cancel a $CNOT$.
In this way, we obtain the explicit circuit for a double qubit excitation, shown in Fig. \ref{fig:d_q_exc_full}. The explicit circuit has a $CNOT$ count of $13$ and $CNOT$ depth of $11$.
The circuit suggested in Refs. \cite{Eff_d_q_exc,Eff_d_q_exc_2}, for the same operation, also has a $CNOT$ count of $13$, but a higher $CNOT$ depth of $13$ 
\footnote{An anonymous referee has pointed out that another advantage of the circuit on Fig. \ref{fig:d_q_exc_full} is that it reduces the number of long-range $CNOT$s compared to the circuits in Refs. \cite{Eff_d_q_exc,Eff_d_q_exc_2}}.

\ 

\onecolumngrid

\twocolumngrid

\subsection{Efficient Fermionic excitations}\label{sec:f_exc}

The quantum-gate operator expressions for fermionic excitations [Eqs. (\ref{eq:s_f_exc_0}) and (\ref{eq:d_f_exc})] differ from those for qubit excitations [Eqs. (\ref{eq:s_exch}) and (\ref{eq:d_q_exc})] only by the additional products of Pauli-$z$ operators in their exponents. These products account for the fermionic anticommutation relations [Eq. \eqref{eq:fermAnti}]: In the single (double) fermionic excitation, the operator product  changes the sign before the excitation parameter $\theta$ if the parity of the set of qubits $\{q_{i+1}..q_{k-1}\}$ ($\{q_{i+1}..q_{j-1}q_{k+1}..q_{l-1}\}$) is odd. 
We can re-express the fermionic excitations in terms of qubit excitations:
\begin{multline}\label{eq:f_exc_1}
U_{ki}^{\mathrm{(sf)}}(\theta) \text{=}
\begin{cases}
U_{ki}^{\mathrm{(sq)}}(\theta) \text{ if } P(q_{i+1}..q_{k-1}) \text{=0}\\
\\
U_{ki}^{\mathrm{(sq)}}(\text{-}\theta) \text{ if } P(q_{i+1}..q_{k-1}) \text{=1}
\end{cases},
\end{multline}
\begin{multline}\label{eq:f_exc_2}
U_{klij}^{\mathrm{(df)}}(\theta) \text{=} 
\begin{cases}
U_{klij}^{\mathrm{(dq)}}(\theta) \text{ if } P(q_{i+1}..q_{j-1}q_{k+1}..q_{l-1})\text{=0}\\
\\
U_{klij}^{\mathrm{(dq)}}(\text{-}\theta) \text{ if } P(q_{i+1}..q_{j-1} q_{k+1}..q_{l-1})\text{=1}
\end{cases}.
\end{multline}
We can adapt the circuits for qubit excitations in Figs. \ref{fig:s_exchange} and \ref{fig:d_q_exc} to perform fermionic excitation, by sandwiching the controlled-$R_y(\theta)$ rotation in each of them between two $CNOT$ staircases [see Figs. \ref{fig:s_fermi} and \ref{fig:d_fermi}].
In this way the sign before the excitation parameter $\theta$ is changed if the parity of the relevant qubits is odd, in accordance with Eqs. (\ref{eq:f_exc_1}) and (\ref{eq:f_exc_2}). Compared to the standard circuits for  fermionic excitations, (see appendix \ref{app:f_circs}), the circuits outlined here utilizes only $2$ $CNOT$ staircases, instead of $4$ or $16$, per single or double fermionic excitation, respectively.

\subsubsection{Circuit for single fermionic excitations}

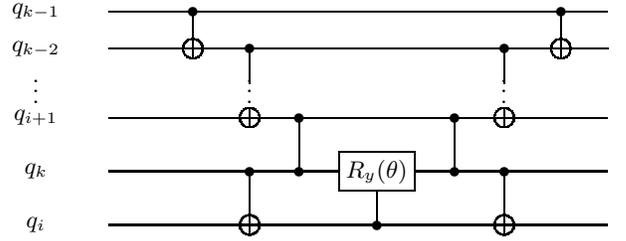
\begin{figure}[H]
\ \ \ \ \ \Qcircuit @C=1.5em @R=1.0em {
q_{k-1} &&& \qw & \ctrl{1} &  \qw & \qw & \qw &\qw & \qw &\ctrl{1}& \qw &\\
q_{k-2} &&& \qw & \targ & \ctrl{1}  & \qw & \qw &\qw &\ctrl{1} &\targ & \qw &\\
\vdots &&&&& \vdots &&&& \vdots &&& \\
q_{i+1}&&& \qw &\qw & \targ &  \ctrl{1} & \qw & \ctrl{1} &  \targ & \qw & \qw &  \\
q_k &&& \qw &\qw & \ctrl{1}  & \ctrl{-1} &  \gate{R_y(\theta)} & \ctrl{-1} &  \ctrl{1} & \qw &\qw \\
q_i&&& \qw &\qw & \targ & \qw   &  \ctrl{-1}  & \qw & \targ{} & \qw & \qw & 
}
\caption{A circuit performing a single fermionic excitation [Eq. (\ref{eq:s_f_exc_0})]. The vertical dots denote $CNOT$ staircases on qubits $q_{i+1}$ to $q_{k-1}$.}
\label{fig:s_fermi}
\end{figure}

Figure \ref{fig:s_fermi} shows our modified circuit for a single fermionic excitation, based on the circuit in Fig. \ref{fig:s_exchange}a.
The parity of $\{q_{i+1}..q_{k-1}\}$ is encoded in qubit $q_{i+1}$  by a $CNOT$ staircase. 
Conditioned on qubit $q_{i-1}$ being in state $|1\rangle$ (odd parity), the two $CZ$ gates between qubits $q_k$ and $q_{i+1}$ reverse the direction of the $R_y(\theta)$ rotation: $R_y(\theta) \rightarrow R_y(-\theta)$.
The controlled $R_y(\theta)$ rotation is implemented as in the single-qubit-excitation circuit, and similarly the circuit identity from Fig. \ref{fig:c_zx} is used to cancel a $CNOT$ between qubits $q_k$ and $q_i$.
A single fermionic excitation, 
Eq. \eqref{eq:s_f_exc_0}, involves operations on qubits $q_i$ to $q_k$.
We define the total number of qubits involved in the excitation as $n^{(\mathrm{sf})} \equiv k -i +1 $. 
For $n^{(\mathrm{sf})} \geq 3$, the circuit in Fig. \ref{fig:s_fermi} has a $CNOT$ count of $2n^{(\mathrm{sf})}-1$ and a $CNOT$ depth of $\mathrm{max}[5,2n^{(\mathrm{sf})}-3]$.
For $n^{(\mathrm{sf})}=2$, the circuit in Fig. \ref{fig:s_fermi} is equivalent to that for a single qubit excitation (Fig. \ref{fig:s_exchange}b).
For comparison, the standard construction (Sec. \ref{sec:stair_case}) for single-fermionic-excitation circuits, yields a circuit (appendix \ref{app:f_circs} Fig. \ref{fig:s_f_exc}) with a $CNOT$ count and a $CNOT$ depth of $4(n^{(\mathrm{sf})} -1)$.
 
\subsubsection{Circuit for double fermionic excitations}

\begin{figure}
\ \ \ \ \ \Qcircuit @C=1.5em @R=1.0em {
q_{l-1} && \qw & \ctrl{1} & \qw & \qw & \qw & \qw & \qw &\ctrl{1}& \qw &\\
q_{l-2} && \qw & \targ & \ctrl{1} & \qw & \qw & \qw &\ctrl{1} &\targ & \qw &\\
\vdots &&&& \vdots &&&& \vdots &&& \\
q_{i+1}&& \qw & \qw & \targ & \ctrl{1} & \qw & \ctrl{1} & \targ & \qw \\
q_l &&\qw & \ctrl{1} & \ctrl{2} & \ctrl{-1} & \gate{R_y(\theta)} & \ctrl{-1} & \ctrl{2}& \ctrl{1} & \qw &\\
q_k && \qw &\targ{}  & \qw & \qw &  \ctrlo{-1} & \qw & \qw &  \targ{} & \qw &\\
q_j  &&\qw & \ctrl{1} &  \targ{} & \qw & \ctrl{-1} & \qw & \targ{} & \ctrl{1} & \qw &\\
q_i &&\qw & \targ{} & \qw & \qw & \ctrlo{-1} & \qw & \qw & \targ{} & \qw &
}
\caption{A circuit performing a double fermionic excitation [Eq.(\ref{eq:d_f_exc})]. The vertical dots denote $CNOT$ staircases on qubits $q_{i+1}$ to $q_{j-1}$, and $q_{k+1}$ to $q_{l-1}$.}
\label{fig:d_fermi}
\end{figure}
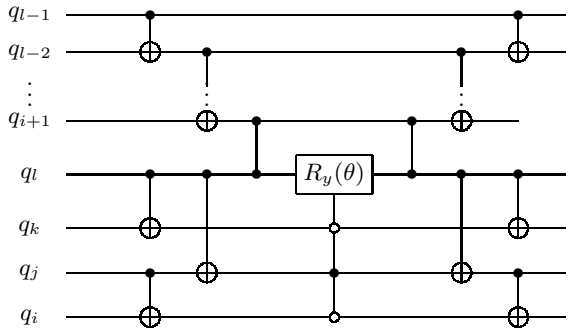

Figure \ref{fig:d_fermi} shows our modified circuit for a double fermionic excitation, based on the circuit in Fig. \ref{fig:d_q_exc}.
The parity of  $\{q_{i+1}..q_{j-1}q_{k+1}..q_{l-1}\}$ is encoded in qubit $q_{i+1}$ by a staircase $CNOT$ structure. 
Conditioned on qubit $q_{i+1}$ being in state $|1\rangle$, the two $CZ$ gates between qubits $q_l$ and $q_{i+1}$  reverse the direction of the controlled $R_y(\theta)$ rotation: $R_y(\theta) \rightarrow R_y(-\theta)$. 
The controlled $R_y(\theta)$ rotation is implemented as in Fig. \ref{fig:cccU1}, and the circuit identity from Fig. \ref{fig:c_zx} is used to cancel a $CNOT$ between qubits $q_l$ and $q_j$ . 
A double fermionic excitations, Eq. \eqref{eq:d_f_exc}, involves operations on qubits $q_i$ to $q_j$ and $q_k$ to $q_l$.
We define  the total number of qubits participating in the excitation as $n^{(\mathrm{df})} \equiv j -i + l -k + 2$. The circuit in Fig. \ref{fig:d_fermi} has a $CNOT$ count of $2n^{(\mathrm{df})} + 5$ and a $CNOT$ depth of $\mathrm{max}[13, 2n^{(\mathrm{df})} - 1]$ for $n^{(\mathrm{df})} \geq 5$. 
For $n^{(\mathrm{df})}=4$, the circuit in Fig. \ref{fig:d_fermi} is equivalent to that for a double qubit excitation (Figs. \ref{fig:d_q_exc} and \ref{fig:d_q_exc_full}).  
For comparison, the standard construction (Sec. \ref{sec:stair_case}) for fermionic-excitation circuits, yields a circuit (appendix \ref{app:f_circs} Fig. \ref{fig:d_f_exc}) with a $CNOT$ count and a $CNOT$ depth of  $16(n^{(\mathrm{df})} -1)$.
Additionaly, using the method of ``balanced trees'', suggested in Ref. \cite{phase_gadget}, one can rearrange the $CNOT$s in the fermionic excitation circuits (Figs. \ref{fig:s_fermi} and \ref{fig:d_fermi}), so that their $CNOT$ depths grow logarithmically rather than linearly.

\section{Summary}\label{summary}

In this work, we constructed circuits, optimized in terms of $CNOT$ gates, to perform efficiently single and double qubit excitations, excitations generated by qubit creation and annihilation operators. We then expanded the circuits' functionality to account for fermionic anticommutation relations, and to perform fermionic excitations. 
Our single-fermionic-excitation circuits use $2 n^{(\mathrm{sf})} -1$ $CNOT$s, and have $CNOT$ depths of $3$ for $n^{(\mathrm{sf})}=2$, and $\mathrm{max}[5,2n^{(\mathrm{sf})}-3]$ for $n^{(\mathrm{sf})} \geq 3$, where $n^{(\mathrm{sf})}$ is the number of orbitals on which the single fermionic excitation acts.
Our double-fermionic-excitation circuits use $2 n^{(\mathrm{df})} +5$ $CNOT$s, and have $CNOT$ depths of $11$ for $n^{(\mathrm{df})}=4$, and $\mathrm{max}[13, 2n^{(\mathrm{df})} - 1]$ for $n^{(\mathrm{df})} \geq 5$, where $n^{(\mathrm{df})}$ is the number of orbitals on which the double fermionic excitation acts. 
To our knowledge, our circuits are more efficient than all previous methods for implementing double qubit and double fermionic excitations---both in $CNOT$ count and in $CNOT$ depth. 
Our optimized circuits present a significant step towards implementations of VQE algorithms on NISQ devices. 
 
\ \\

\begin{acknowledgements}
	The authors wish  to thank V. Armaos for useful discussions, and an anonymous referee for pointing out the advantage of our methods mentioned in Footnote 2. Y.S.Y. acknowledges financial support from the EPSRC and Hitachi via CASE studentship RG97399. D.R.M.A.-S. was supported by the EPSRC, Lars Hierta's Memorial Foundation and Girton College. 
\end{acknowledgements}
\ \\

\appendix

\section{Standard single and double fermionic excitations circuits}\label{app:f_circs}

\begin{figure}[H]
\ \ \ \ \ \Qcircuit @C=0.9em @R=0.7em {
q_k && \gate{R_x(\frac{\pi}{2})} & \ctrl{1} & \qw & \qw &  \qw &\qw & \qw &\ctrl{1}& \gate{R_x(-\frac{\pi}{2})} & \qw & \gate{H} & \ctrl{1} & \qw & \qw &  \qw &\qw & \qw &\ctrl{1}& \gate{H} & \qw \\
q_{k-1} && \qw & \targ & \ctrl{1} & \qw & \qw & \qw &\ctrl{1} &\targ & \qw & \qw  & \qw & \targ & \ctrl{1} & \qw & \qw & \qw &\ctrl{1} &\targ & \qw & \qw \\
\vdots &&&& \vdots &&&& \vdots &&&  &&& \vdots &&&& \vdots &&& \\
q_{i+1} && \qw & \qw & \targ &  \ctrl{1} & \qw & \ctrl{1}  & \targ & \qw & \qw & \qw & \qw & \qw & \targ &  \ctrl{1} & \qw & \ctrl{1}  & \targ & \qw & \qw & \qw \\
q_{i} && \gate{H}  &\qw & \qw & \targ & \gate{R_z(\theta)} & \targ & \qw & \qw & \gate{H} & \qw  & \gate{R_x(\frac{\pi}{2})}  &\qw & \qw & \targ & \gate{R_z(-\theta)} & \targ & \qw & \qw & \gate{R_x(-\frac{\pi}{2})} & \qw \\
}
\caption{A standard circuit performing a single fermionic excitation [Eq. \eqref{eq:s_f_exc_0}]. The vertical dots denote $CNOT$ staircases on qubits $q_{i+1}$ to $q_{k-1}$.}
\label{fig:s_f_exc}
\end{figure}
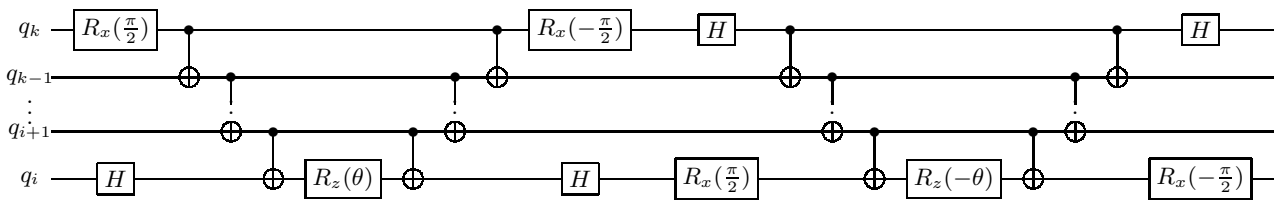

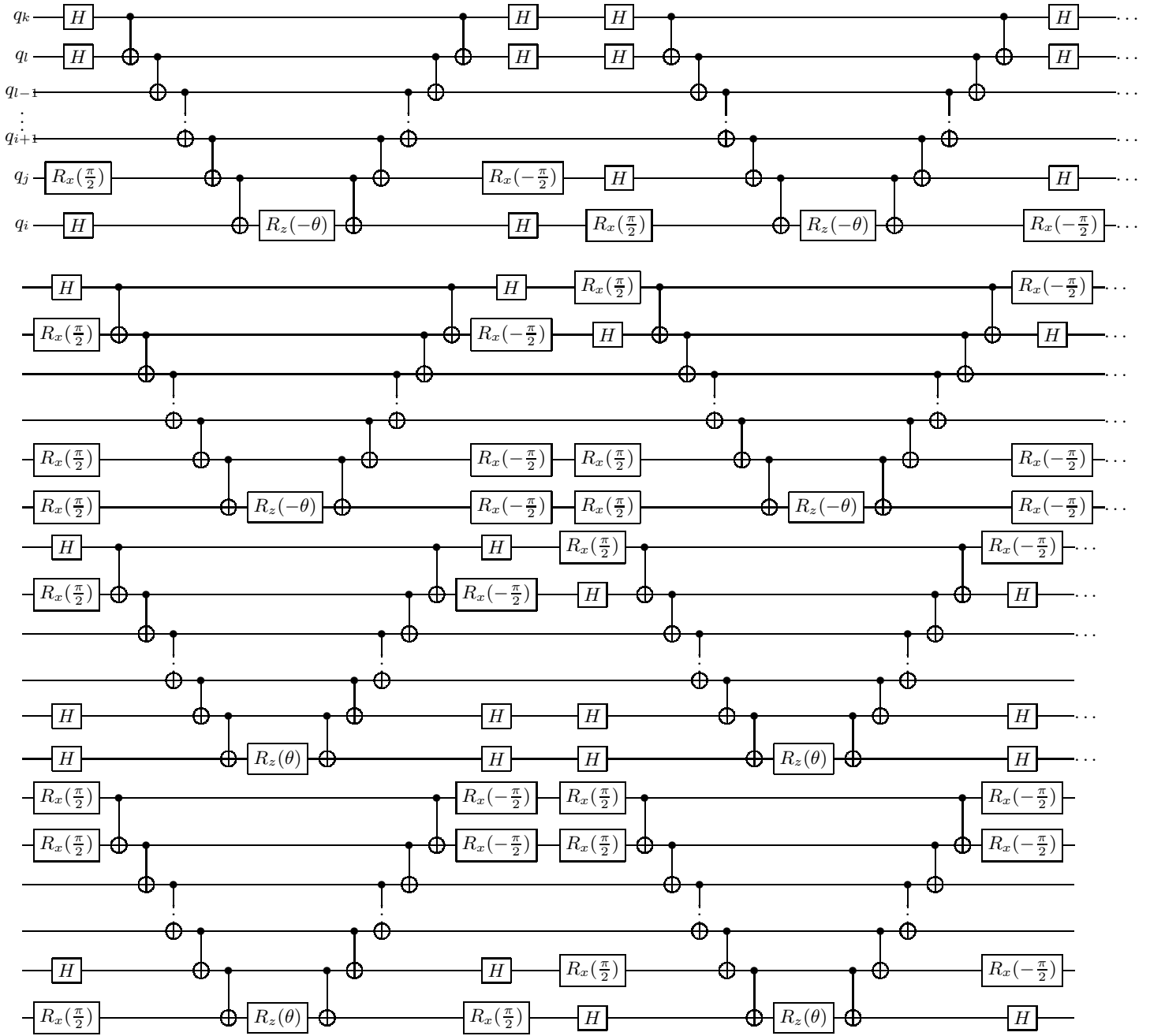
\begin{figure}[H]
\Qcircuit @C=0.6em @R=0.8em {
q_k && \gate{H} & \ctrl{1} & \qw & \qw & \qw & \qw & \qw & \qw &  \qw &\qw & \qw &\ctrl{1}& \gate{H} & \qw           & \gate{H} & \ctrl{1} & \qw & \qw & \qw & \qw & \qw & \qw &  \qw &\qw & \qw &\ctrl{1}& \gate{H} & \qw  & \hdots \\
q_{l} && \gate{H} & \targ & \ctrl{1} & \qw & \qw & \qw & \qw & \qw & \qw & \qw &\ctrl{1} &\targ & \gate{H} & \qw            & \gate{H} & \targ & \ctrl{1} & \qw & \qw & \qw & \qw & \qw & \qw & \qw &\ctrl{1} &\targ & \gate{H} & \qw & \hdots \\
q_{l-1} && \qw & \qw & \targ & \ctrl{1} & \qw & \qw & \qw & \qw & \qw & \ctrl{1} &\targ &\qw & \qw & \qw                 & \qw & \qw & \targ & \ctrl{1} & \qw & \qw & \qw & \qw & \qw & \ctrl{1} &\targ &\qw & \qw & \qw & \hdots \\
\vdots &&&&& \vdots &&&&&& \vdots &&&             &&&&& \vdots &&&&&& \vdots &&& \\
q_{i+1} && \qw &  \qw & \qw & \targ & \ctrl{1} &  \qw & \qw & \qw  & \ctrl{1} & \targ & \qw & \qw & \qw & \qw             & \qw &  \qw & \qw & \targ & \ctrl{1} &  \qw & \qw & \qw  & \ctrl{1} & \targ & \qw & \qw & \qw & \qw & \hdots \\
q_{j} && \gate{R_x(\frac{\pi}{2})} &  \qw & \qw & \qw & \targ &  \ctrl{1} & \qw & \ctrl{1}  & \targ & \qw & \qw & \qw & \gate{R_x(-\frac{\pi}{2})} & \qw            & \gate{H} &  \qw & \qw & \qw & \targ &  \ctrl{1} & \qw & \ctrl{1}  & \targ & \qw & \qw & \qw & \gate{H} & \qw & \hdots \\
q_{i} && \gate{H}  & \qw & \qw & \qw & \qw & \targ & \gate{R_z(-\theta)} & \targ & \qw & \qw & \qw & \qw & \gate{H} & \qw                  & \gate{R_x(\frac{\pi}{2})}  & \qw & \qw & \qw & \qw & \targ & \gate{R_z(-\theta)} & \targ & \qw & \qw & \qw & \qw & \gate{R_x(-\frac{\pi}{2})} & \qw & \hdots \\
}
\ \\
\ \\
\Qcircuit @C=0.6em @R=0.8em {
& \gate{H} & \ctrl{1} & \qw & \qw & \qw & \qw & \qw & \qw &  \qw &\qw & \qw &\ctrl{1}& \gate{H} & \qw           & \gate{R_x(\frac{\pi}{2})}  & \ctrl{1} & \qw & \qw & \qw & \qw & \qw & \qw &  \qw &\qw & \qw &\ctrl{1}& \gate{R_x(-\frac{\pi}{2})}  & \qw  & \hdots \\
& \gate{R_x(\frac{\pi}{2})}  & \targ & \ctrl{1} & \qw & \qw & \qw & \qw & \qw & \qw & \qw &\ctrl{1} &\targ & \gate{R_x(-\frac{\pi}{2})} & \qw            & \gate{H} & \targ & \ctrl{1} & \qw & \qw & \qw & \qw & \qw & \qw & \qw &\ctrl{1} &\targ & \gate{H} & \qw & \hdots \\
& \qw & \qw & \targ & \ctrl{1} & \qw & \qw & \qw & \qw & \qw & \ctrl{1} &\targ &\qw & \qw & \qw                 & \qw & \qw & \targ & \ctrl{1} & \qw & \qw & \qw & \qw & \qw & \ctrl{1} &\targ &\qw & \qw & \qw & \hdots \\
&&&& \vdots &&&&&& \vdots &&&&             &&&& \vdots &&&&&& \vdots &&& \\
& \qw &  \qw & \qw & \targ & \ctrl{1} &  \qw & \qw & \qw  & \ctrl{1} & \targ & \qw & \qw & \qw & \qw             & \qw &  \qw & \qw & \targ & \ctrl{1} &  \qw & \qw & \qw  & \ctrl{1} & \targ & \qw & \qw & \qw & \qw & \hdots \\
& \gate{R_x(\frac{\pi}{2})} &  \qw & \qw & \qw & \targ &  \ctrl{1} & \qw & \ctrl{1}  & \targ & \qw & \qw & \qw & \gate{R_x(-\frac{\pi}{2})} & \qw            & \gate{R_x(\frac{\pi}{2})}  &  \qw & \qw & \qw & \targ &  \ctrl{1} & \qw & \ctrl{1}  & \targ & \qw & \qw & \qw & \gate{R_x(-\frac{\pi}{2})} & \qw & \hdots \\
& \gate{R_x(\frac{\pi}{2})}   & \qw & \qw & \qw & \qw & \targ & \gate{R_z(-\theta)} & \targ & \qw & \qw & \qw & \qw & \gate{R_x(-\frac{\pi}{2})}  & \qw                  & \gate{R_x(\frac{\pi}{2})}  & \qw & \qw & \qw & \qw & \targ & \gate{R_z(-\theta)} & \targ & \qw & \qw & \qw & \qw & \gate{R_x(-\frac{\pi}{2})} & \qw & \hdots \\
}
\ \\
\ \\
\Qcircuit @C=0.6em @R=0.8em {
& \gate{H} & \ctrl{1} & \qw & \qw & \qw & \qw & \qw & \qw &  \qw &\qw & \qw &\ctrl{1}& \gate{H} & \qw           & \gate{R_x(\frac{\pi}{2})} & \ctrl{1} & \qw & \qw & \qw & \qw & \qw & \qw &  \qw &\qw & \qw &\ctrl{1}& \gate{R_x(-\frac{\pi}{2})} & \qw  & \hdots \\
& \gate{R_x(\frac{\pi}{2})} & \targ & \ctrl{1} & \qw & \qw & \qw & \qw & \qw & \qw & \qw &\ctrl{1} &\targ & \gate{R_x(-\frac{\pi}{2})} & \qw            & \gate{H} & \targ & \ctrl{1} & \qw & \qw & \qw & \qw & \qw & \qw & \qw &\ctrl{1} &\targ & \gate{H} & \qw & \hdots \\
& \qw & \qw & \targ & \ctrl{1} & \qw & \qw & \qw & \qw & \qw & \ctrl{1} &\targ &\qw & \qw & \qw                 & \qw & \qw & \targ & \ctrl{1} & \qw & \qw & \qw & \qw & \qw & \ctrl{1} &\targ &\qw & \qw & \qw & \hdots \\
&&&& \vdots &&&&&& \vdots &&&&             &&&& \vdots &&&&&& \vdots &&& \\
& \qw &  \qw & \qw & \targ & \ctrl{1} &  \qw & \qw & \qw  & \ctrl{1} & \targ & \qw & \qw & \qw & \qw             & \qw &  \qw & \qw & \targ & \ctrl{1} &  \qw & \qw & \qw  & \ctrl{1} & \targ & \qw & \qw & \qw & \qw &  \\
& \gate{H} &  \qw & \qw & \qw & \targ &  \ctrl{1} & \qw & \ctrl{1}  & \targ & \qw & \qw & \qw & \gate{H} & \qw            & \gate{H} &  \qw & \qw & \qw & \targ &  \ctrl{1} & \qw & \ctrl{1}  & \targ & \qw & \qw & \qw & \gate{H} & \qw & \hdots \\
& \gate{H}  & \qw & \qw & \qw & \qw & \targ & \gate{R_z(\theta)} & \targ & \qw & \qw & \qw & \qw & \gate{H} & \qw                  & \gate{H}  & \qw & \qw & \qw & \qw & \targ & \gate{R_z(\theta)} & \targ & \qw & \qw & \qw & \qw & \gate{H} & \qw & \hdots \\
}
\ \\
\ \\
\Qcircuit @C=0.6em @R=0.8em {
& \gate{R_x(\frac{\pi}{2})} & \ctrl{1} & \qw & \qw & \qw & \qw & \qw & \qw &  \qw &\qw & \qw &\ctrl{1}& \gate{R_x(-\frac{\pi}{2})} & \qw           & \gate{R_x(\frac{\pi}{2})}  & \ctrl{1} & \qw & \qw & \qw & \qw & \qw & \qw &  \qw &\qw & \qw &\ctrl{1}& \gate{R_x(-\frac{\pi}{2})}  & \qw  \\
& \gate{R_x(\frac{\pi}{2})}  & \targ & \ctrl{1} & \qw & \qw & \qw & \qw & \qw & \qw & \qw &\ctrl{1} &\targ & \gate{R_x(-\frac{\pi}{2})} & \qw            & \gate{R_x(\frac{\pi}{2})} & \targ & \ctrl{1} & \qw & \qw & \qw & \qw & \qw & \qw & \qw &\ctrl{1} &\targ & \gate{R_x(-\frac{\pi}{2})} & \qw  \\
& \qw & \qw & \targ & \ctrl{1} & \qw & \qw & \qw & \qw & \qw & \ctrl{1} &\targ &\qw & \qw & \qw                 & \qw & \qw & \targ & \ctrl{1} & \qw & \qw & \qw & \qw & \qw & \ctrl{1} &\targ &\qw & \qw & \qw  \\
&&&& \vdots &&&&&& \vdots &&&&             &&&& \vdots &&&&&& \vdots &&& \\
& \qw &  \qw & \qw & \targ & \ctrl{1} &  \qw & \qw & \qw  & \ctrl{1} & \targ & \qw & \qw & \qw & \qw             & \qw &  \qw & \qw & \targ & \ctrl{1} &  \qw & \qw & \qw  & \ctrl{1} & \targ & \qw & \qw & \qw & \qw  \\
& \gate{H} &  \qw & \qw & \qw & \targ &  \ctrl{1} & \qw & \ctrl{1}  & \targ & \qw & \qw & \qw & \gate{H} & \qw            & \gate{R_x(\frac{\pi}{2})}  &  \qw & \qw & \qw & \targ &  \ctrl{1} & \qw & \ctrl{1}  & \targ & \qw & \qw & \qw & \gate{R_x(-\frac{\pi}{2})} & \qw  \\
& \gate{R_x(\frac{\pi}{2})}   & \qw & \qw & \qw & \qw & \targ & \gate{R_z(\theta)} & \targ & \qw & \qw & \qw & \qw & \gate{R_x(\frac{\pi}{2})}  & \qw                  & \gate{H}  & \qw & \qw & \qw & \qw & \targ & \gate{R_z(\theta)} & \targ & \qw & \qw & \qw & \qw & \gate{H} & \qw  \\
}
\caption{A standard circuit performing a double fermionic excitation [Eq. \eqref{eq:d_f_exc}]. The vertical dots denote $CNOT$ staircases on qubits $q_{i+1}$ to $q_{j-1}$, and $q_{k+1}$ to $q_{l-1}$.}
\label{fig:d_f_exc}
\end{figure}

In this appendix, we present the full circuits for single and double fermionic excitations constructed with the standard use of $CNOT$ staircases.
As described in Sec. \ref{sec:stair_case}, a single fermionic excitation [Eq. \eqref{eq:s_f_exc_0}] can be implemented by the circuit in Fig. \ref{fig:s_f_exc}, which contains $2$ $CNOT$ staircase constructions.
Similarly, a double fermionic excitations [Eq. \eqref{eq:d_f_exc}] can be implemented by the circuit in Fig. \ref{fig:d_f_exc}, which contains $8$ $CNOT$ staircase constructions.

\onecolumngrid

\twocolumngrid

\bibliographystyle{apsrev4-1}
\bibliography{references}

\end{document}